\documentclass[twocolumn,showpacs,preprintnumbers,amsmath,amssymb]{revtex4}
\usepackage{graphicx}
\usepackage{dcolumn}
\usepackage{bm}
\newcommand{\be}{\begin{equation}}
\newcommand{\ee}{\end{equation}}
\newcommand{\bef}{\begin{figure}}
\newcommand{\eef}{\end{figure}}
\newcommand{\bea}{\begin{eqnarray}}
\newcommand{\eea}{\end{eqnarray}}
\begin{document}
\preprint{APS/123-QED}

\title{Scale invariant forces in $1d$ shuffled lattices}

\author{Andrea Gabrielli$^{1,2}$}
\affiliation{$^1$ Istituto dei Sistemi Complessi - CNR, Via dei Taurini 19, 
00185-Rome (Italy)}
\affiliation{$^2$ SMC-INFM, Department of Physics, University 
``La Sapienza'' of Rome, P.le Aldo Moro 2, 00185-Rome, Italy}
\date{\today}

\hyphenation{Lan-ge-vin}
\begin{abstract}
In this paper we present a detailed and exact study of the probability
density function $P(F)$ of the total force $F$ acting on a point
particle belonging to a perturbed lattice of identical point
sources of a power law pair interaction. 
The main results concern the large $F$ tail of
$P(F)$ for which two cases are mainly distinguished: (i) Gaussian-like
fast decreasing $P(F)$ for lattice with perturbations forbidding any
pair of particles to be found arbitrarily close to one each other;
(ii) L\'evy-like power law decreasing $P(F)$ when this possibility is
instead permitted.  It is important to note that in the second case the
exponent of the power law tail of $P(F)$ is the same for all
perturbation (apart from very singular cases), and is in an one to one
correspondence with the exponent characterizing the behavior of the
pair interaction with the distance between the two particles.
\end{abstract}

\maketitle

\section{Introduction}
The knowledge of the statistical properties of the force acting on a
particle belonging to a gas and exerted by all the other particles provides
important information in many physical contexts and applications.
Typical examples are: (i) distribution of the gravitational force in a
gas of masses in cosmological and stellar astrophysical applications
\cite{chandra,star,fractal}, (ii) distribution of molecular and
dipolar interactions \cite{dipole} in gas of particles, (iii) theory
of defects in condensed matter physics \cite{dislo}, and (iv) granular
materials \cite{granular1,granular2}.  The first seminal work in this
field was due to Chandrasekhar \cite{chandra} and deals, among many
subjects, with the gravitational force probability distribution in a
homogeneous Poisson spatial distribution of identical particles. By
studying the characteristic function of the sum of the stochastic
forces due to the single particles, the probability distribution of
this total force is exactly found to be given by the so-called Holtzmark
distribution, which is a three-dimensional analog of the
one-dimensional fat tailed stable L\'evy distributions.  In
\cite{star,gauss-poisson,dipole,libro} approximated extensions, to
different branches of physics, of this approach can be found for more
complex particle distribution (i.e., point processes) obtained by
perturbing a homogeneous Poisson point process.  In this paper we
present a study of the total force probability distribution for a very
different class of spatial particle distributions (i.e., point
processes), the perturbed lattices of point particles, in the case in
which the pair interaction decays spatially as a general power law. We
think that this study can be very useful for application in both solid
state physics (e.g., in the case of Coulomb or dipolar pair
interaction) \cite{dipole}, and cosmology where $n$-body
gravitational simulations (introduced to study the problem of
``structure formation'' due to gravitational collapse from primordial 
cosmological mass density fluctuations) 
are performed usually starting from suitable perturbed
lattice initial conditions \cite{n-body}.  We limit the study to the
one-dimensional case in order to avoid difficulties related to the
anisotropies of higher dimensional lattices. However the exact results
we present in $1d$ are suggestive of the behavior of the same quantities in
higher dimensions. In fact one can see \cite{3d-ours} that the change of
spatial dimension only renders calculation not explicitly performable,
keeping qualitatively the behavior we present below.

\section{Definitions and formalism}

In order to approach in the proper way the problem of the global force
probability distribution in a perturbed lattice of particles
interacting via a power law spatially decreasing pair interaction, let
us consider firstly a gas of such identical particles with microscopic
density
\[n(x)=\sum_i\delta(x-x_i)\,,\]
where $x_i$ is the position of the $i^{th}$ particle. 
Let us assume that the average number density $n_0=\left<n(x)\right>>0$
(where the average $\left<...\right>$ is to be intended as an ensemble 
average) is well defined (i.e., the particle distribution is 
uniform on sufficiently large scales \cite{libro}). 
We then suppose that particles interact via a pair force $f(x)$ 
depending on the mutual pair distance $x$ as
\[f(x)=-C\frac{x}{|x|^{\alpha+1}}\,.\]
This means that $f(x)$ gives the force exerted by a particle in the origin 
on another particle
in $x$ (the force is attractive if $C>0$, and repulsive if $C<0$).
Therefore the force-field in the point $x$ of the space will be
\be
{\cal F}(x)=C\sum_i\frac{x_i-x}{|x_i-x|^{\alpha+1}}\equiv
C\int dy\,n(y)\frac{y-x}{|y-x|^{\alpha+1}}\,,
\label{pre1}
\ee
where the last
integral is over all the space.  However, from the last expression of
Eq.~(\ref{pre1}), being $n_0>0$, we have that for a given
realization of the stochastic density field $n(x)$, the infinite
volume limit of ${\cal F}(x)$ is not univocally defined for $\alpha\le
1$ (i.e. the integral in {\em absolutely diverging}, and 
its value depend on how this limit is taken). The same
feature is present in higher spatial dimensions. For example in $d=3$
the same problem is present for $\alpha\le 3$ and in general
$\alpha\le d$ in $d$ dimensions. For instance this is the case of the
gravitational force in a self-gravitating homogeneous gas of identical
masses \cite{chandra}, and of Coulomb interaction in the one component
plasma (OCP) \cite{OCP} of identical electrical charges both in the
disordered and the ordered (i.e., the Coulomb lattice \cite{pines})
phases.  This problem is well known in condensed matter physics about
the OCP.  However, in this case, the problem is automatically solved
by the presence in the physical system of a uniform background charge
density $n_b(x)=-n_0$ with opposite sign with respect to the identical
charged particles and such that to conserve global charge neutrality
in the system.  Once the attractive force of the background is
considered on a charged particle together with the repulsive forces
exerted by the other particles, the problem of the infinite volume
limit of ${\cal F}$ is solved and its value unique.  For what concerns
the self-gravitating systems, in Newtonian gravitation an analog of
the uniform background of the OCP (i.e. a negative uniform mass
density $n_b(x)=-n_0$ such to generate a repulsive force on the particles) 
does not exist, and it has to be introduced in
the system artificially to regularize the problem (an approach usually
called {\em Jeans' swindle} \cite{binney-tremaine}). However this negative
background comes out
naturally, as an effect of space expansion, when the gravitational
motion of particles is described, starting from the equations of
general relativity, in comoving coordinates in a quasi-uniform
expanding Einstein - De Sitter universe \cite{peebles80} which is the
main model of universe used in cosmology.  
In practice considering the presence of such balancing background
will give for ${\cal F}$ the following expression:
\be
{\cal F}(x)=C\int dy\,\delta n(y)\frac{y-x}{|y-x|^{\alpha+1}}\,,
\label{pre2}
\ee where $\delta n(x)=n(x)-n_0$. This makes ${\cal F}$ to be defined
also for smaller $\alpha$ depending on the small $k$ behavior of the power
spectrum $S(k)\sim \left<|\tilde{\delta n}(k)|^2\right>$ of the
density field where $\tilde{\delta n}(k)$ is the Fourier transform of
$\delta n(x)$.  By studying the large distance scaling behavior of the
integrated fluctuations of $n(x)$ \cite{libro}, it is simple to show
that, assuming $S(k)\sim k^\beta$ at small $k$, in $d$ dimension
${\cal F}$ is a well defined statistical quantity (i.e., its value
does not depend on the way in which the infinite volume limit is
taken) for $\alpha>(d-\beta)/2$ if $\beta<1$, and $\alpha>(d-1)/2$ if
$\beta\ge 1$ (see also the discussion in Appendix II on the
definiteness of the force ${\cal F}$ respectively in the shuffled
lattice and the homogeneous Poisson particle distributions).  Note
finally that taking the infinite volume limit symmetrically with
respect to the point $x$ on which the force is calculated the
background gives a zero net force on the point $x$. Therefore the
value of ${\cal F}$ obtained calculating Eq.~(\ref{pre1}) taking the
infinite volume limit symmetrically with respect to the point $x$
gives automatically the well defined value obtained by
Eq.~(\ref{pre2}) (i.e., subtracting the effect of the background).
This preliminary discussion of the statistical definiteness of the
force ${\cal F}$ is useful to justify the symmetrical way in which we
take the infinite volume limit in the one-dimensional shuffled lattice
case we analyze in the rest of the paper.  The statistical properties
of ${\cal F}$ we will find in this peculiar way (no background and
symmetrical limit) coincide with those of the case in which the effect
of a negative background is considered independently of the way in
which the infinite volume limit is taken, which therefore can be
considered the {\em real} physical case. Moreover, from the above
considerations we deduce that, as in the shuffled lattice $S(k)\sim
k^2$ at small $k$ (see \cite{displa}), the results are valid for all
values $\alpha>0$ in $d=1$.


Let us take, therefore, a $1d$ regular chain of $2N+1$ unitary mass
particles with a lattice spacing $a>0$ (we will take eventually the
limit $N\rightarrow +\infty$), i.e., the position of the $n^{th}$
particle is $X_n=na$. Therefore the microscopic density can be written
as
\[
n_{in}(x)=\sum_{n=-N}^N \delta(x-na)\,.
\]

Clearly the average density of particles in the system is $n_0=1/a$.
We now apply an uncorrelated displacement field (i.e., a random
shuffling) to this system, i.e., a random
displacement $U_n$ is applied to the generic $n^{th}$ particle 
independently of the other particles.
This displacement field is completely characterized by the
one-displacement probability density function (PDF) $p(u)$ (i.e.,
$Prob(u\le U_n< u+du)=p(u)du$).  After the application of the
displacements the new microscopic density will be:
\be
n(x)=\sum_{n=-N}^N \delta(x-na-u_n)\,,
\label{eq2}
\ee
the $u_n$'s being the realizations of the random variables $U_n$
all extracted from $p(u)$ independently one of each other. 
We consider the case $p(u)=p(-u)$ for
simplicity.  Equation (\ref{eq2}) says that the particle originally in
$X_n=na$, after the displacement will be in $X_n=na+U_n$. 
For the analysis of spatial density correlations
in such a system see \cite{displa}.
Let us call $q_n(x)$ the PDF of the position of the $n^{th}$ particle
(i.e., $Prob(x\le x_n <x+dx)=q_n(x)dx$).
Clearly it is given by
\[q_n(x)\equiv p(x-na)\,.\]

Let us now assume, as above, that the $n^{th}$ particle creates a force-field 
in the point $x$ of the type: 
\[ f_n(x)=C\frac{X_n-x}{|X_n-x|^{\alpha+1}}\] 
with $\alpha>0$ and $C$ a constant.  Therefore the total stochastic field 
${\cal F}$ generated at a generic point $x$ of
the space by all the system particles is: 
\be 
{\cal F}(x)=C\sum_{n=-N}^{N}\frac{X_n-x}{|X_n-x|^{\alpha+1}}\,.
\label{eq0a}
\ee
Note that it is a sum of random variables.
Let us call $W_0(F)$ its PDF; it will be given by
\bea
W_0(F)=&&\int\int_{-\infty}^{\infty}\left[\prod_{n=-N}^{N}
dx_n\, p(x_n-na)\right]\nonumber\\
&&\times\delta\left(F- C\sum_{n\ne 0}^{-N,N}
\frac{x_n-x}{|x_n-x|^{\alpha+1}}\right)\,.
\nonumber
\eea 
It is immediate to see that ${\cal F}(x)$ is the sum of
independent random variables $C(X_n-x)/|X_n-x|^{\alpha+1}$. However,
as the PDF's $q_n(x)$ change with $n$, these variable are not
identically distributed.  As shown below, this, together with the fact
that ${\cal F}$ needs not a normalization in $N$ to be well defined in
the large $N$ limit, are the reasons why we do not obtain in general an
exact Gaussian or L\'evy limit \cite{levy-stable,feller} for the $W_0(F)$.  
In order to study the asymptotic behavior in $F$ and $N$, 
it is usual to introduce the so-called {\em
characteristic function} of ${\cal F}$, i.e., the Fourier transform
(FT) of $W_0(F)$: 
\bea 
&&\hat W_0(k)\equiv\int_{-\infty}^{\infty} dF\,
W_0(F)e^{ikF}\nonumber\\ 
&&=\prod_{n=-N}^{N}
\int_{-\infty}^{\infty}dy\,p(y-na)\exp\left(iCk\frac{y-x}
{|y-x|^{\alpha+1}}\right)\,.
\label{eq0c}
\eea 
By studying the small $k$ behavior of the single integrals in
Eq.~(\ref{eq0c}) and taking appropriately the limit
$N\rightarrow\infty$ we can deduce the moments and the large $F$ behavior
of $W_0(F)$.  

However, we will study a slightly different and more difficult
problem, which is of particular interest if we want to study the
dynamics of the system particles under the effect of only this mutual
force.  We study directly the statistical properties of the stochastic
force acting on one generic system particle. In particular we
calculate the total force acting on the particle initially located at
the origin of the space and displaced to $X_0=U_0$: \be {\cal
F}=\sum_{n\ne 0}^{-N,N} f_n(X_0)=C\sum_{n\ne 0}^{-N,N}
\frac{X_n-X_0}{|X_n-X_0|^{\alpha+1}}\,.
\label{eq4}
\ee In this way, taking the limit $N\rightarrow\infty$, we get the
symmetric infinite volume limit with no negative background of
Eq.(\ref{pre1}) with $x=X_0$, which is, as explained in the first
paragraph of this section, equal to Eq.(\ref{pre2}) with arbitrary way
of taking the infinite volume limit, and therefore giving the force on
any particle belonging to the system with the uniform balancing
background.  Note that now, because of the presence of the variable
$X_0$ in each term of the sum (\ref{eq4}), ${\cal F}$ is no more a sum
of independent terms.  However we show below how to reduce the problem
to that of a sum of independent stochastic terms, by introducing the
concept of {\em conditional probability density function}.  The
solution of this {\em conditional} problem will give also the way to
face the study of the first {\em unconditional} case given by
Eq.~(\ref{eq0a}).

About ${\cal F}$ in Eq.~(\ref{eq4}), we want again to find the PDF
$W(F)$ of the value $F$ of this force.  As before, since the displacements
applied to the particles are independent one of each other, we have the
exact relation: 
\bea
&&W(F)=\int\int_{-\infty}^{+\infty}
\left[\prod_{n=-N}^{N}dx_n\, p(x_n-na)\right]\nonumber\\
&&\times\delta\left(F- C\sum_{n\ne
0}^{-N,N} \frac{x_n-x_0}{|x_n-x_0|^{\alpha+1}}\right)\,,
\nonumber
\eea 
that, through a simple change of variables $\Delta_n=x_n-x_0$ or $n\ne
0$, can be rewritten as
\bea
\nonumber
&&W(F)=\int_{-\infty}^{+\infty}dx_0 \,p(x_0)\\
&&\int\int_{-\infty}^{+\infty}
\left[\prod_{n\ne 0}^{-N,N}d\Delta_n\,
p(\Delta_n+x_0-na)\right]\nonumber\\
&&\times\delta\left(F- C\sum_{n\ne 0}^{-N,N}
\frac{\Delta_n}{|\Delta_n|^{\alpha+1}}\right)\,.
\nonumber
\eea 
Let us analyze the behavior of the conditional PDF $P(F;x_0)$,
conditioned to the fact that the particle on which the force is 
evaluated is at $X_0=x_0$:
\bea
&&P(F;x_0)=\int\int_{-\infty}^{+\infty}
\left[\prod_{n\ne 0}^{-N,N}d\Delta_n\,
p(\Delta_n+x_0-na)\right]\nonumber\\
&&\times\delta\left(F- C\sum_{n\ne 0}^{-N,N}
\frac{\Delta_n}{|\Delta_n|^{\alpha+1}}\right)\,.
\label{eq6}
\eea 
In this way, once $x_0$ is fixed, the total force ${\cal F}$ is the sum 
of independent contributions
\be
f_n=C\frac{\Delta_n}{|\Delta_n|^{\alpha+1}}\,.
\label{eq6b}
\ee
It is interesting to see in which cases ${\cal F}$ satisfies the central limit
theorem. We will see that it never satisfies this theorem even when
its PDF is rapidly decreasing at large values.  More precisely, we
will see that even in this last case its PDF is dependent on
the details of $p(u)$.

For the sake of simplicity of notation let us assume in the rest of the
paper that $C=1$ (the repulsive case $C=-1$ will be trivially deduced from this).
By performing the simple change of variables given by Eq.~(\ref{eq6b}),
it is possible to find the conditional (i.e., conditioned to $X_0=x_0$)
PDF $g_n(f;x_0)$ of the single stochastic force generated by the 
particle in $X_n$ on the particle fixed in $X_0=x_0$:
\be
g_n(f;x_0)=\frac{|f|^{-1-1/\alpha}}{\alpha}\cdot 
p\left(\frac{f}{|f|^{1+1/\alpha}}+x_0-na\right)\,.
\label{gn}
\ee 
Clearly also the forces $f_n$ so distributed are independent of one
each other. The support of $g_n(f;x_0)$ can be simply deduced from the
one of $p(u)$.

\section{Large $F$ analysis and limit theorems}

Let us take the FT of Eq.~(\ref{eq6}) in order to
evaluate the conditional {\em characteristic function} of the
stochastic variable ${\cal F}$:
\bea
\label{eq7}
&&\hat P(k;x_0)\equiv\int_{-\infty}^{+\infty}dF\,P(F;x_0)\,e^{ikF}\\
&&= \prod_{n\ne 0}^{-N,N}\int_{-\infty}^{+\infty}dx\, p(x+x_0-na) \exp
\left(ik\frac{x}{|x|^{\alpha+1}}\right)\,.\nonumber 
\eea 
Note that the quantity
\bea
\label{eq8b}
&&\int_{-\infty}^{+\infty}dx\, p(x+x_0-na) \exp
\left(ik\frac{x}{|x|^{\alpha+1}}\right)\\
&&=\int_{-\infty}^{+\infty}df\,
g_n(f;x_0)e^{ikf}\equiv \{e^{ikf}\}_{n;x_0}\,, 
\nonumber
\eea 
where $g_n(f;x_0)$,
given by Eq.~(\ref{gn}), is the conditional PDF of the field $f$ felt
by the particle in $x_0$ due to the particle in $x_n$, is the
conditional characteristic function of this field $f$. Moreover
$\{a(f)\}_{n;x_0}=\int_{-\infty}^{+\infty}df\,g_n(f;x_0)a(f)$.  It is
important to note that, if $x_0=0$ (i.e., the particle on which we
calculate the force is stuck at the origin), the condition
$p(u)=p(-u)$ on the displacements of any particle would imply
\bea
&&\int_{-\infty}^{+\infty}dx\,p(x+na) \exp \left(
ik\frac{x}{|x|^{\alpha+1}}\right)\nonumber\\
&&=\int_{-\infty}^{+\infty} dx\,p(x-na)
\exp \left(-ik\frac{x}{|x|^{\alpha+1}}\right)\,.
\nonumber
\eea 
Consequently, if one fixes $x_0=0$ Eq.~(\ref{eq7}) becomes:
\be \hat W(k)= \prod_{n=1}^{N}\left|\int_{-\infty}^{+\infty} dx\,
p(x-na) \exp \left(ik\frac{x}{|x|^{\alpha+1}}\right)\right|^2\,.
\nonumber
\ee 
However the shift $x_0\ne 0$ of the particle initially at the
origin, and on which we calculate the force, breaks this symmetry,
which is anyway recovered when the average over $p(x_0)$ is performed
to proceed from $P(F;x_0)$ to $W(F)$ (however we will see that this is
a further source of noise when we calculate explicitly the variance
of the force ${\cal F}$).

In order to proceed into the analysis of the PDF of ${\cal F}$, we have to
distinguish two basically different cases:
\begin{enumerate}
\item {\bf Non-Overlapping condition} (NOC): 
No particle can be found arbitrarily close to any other particle; i.e., 
the supports\footnote{We call {\em support} of $p(u)$ simply the set
of real values of $u$ such that $p(u)>0$.} respectively of $p(u)$ and
of $p(u-na)$, for all integer $n\ne 0$, have an empty overlap. The main 
case of physical interest in this class of displacement fields
is when $\exists 0<u_0<a/2$ such that $p(u)=0$ for $|u|>u_0$;
\item {\bf Overlapping condition} (OC): 
Particles can cross one each other and at least one pair of particles
can be found arbitrarily close to one each other; i.e., the supports
respectively of $p(u)$ and of $p(u-na)$, for at least an integer $n\ne
0$, have a non-zero overlap.  The main case of physical interest in which
this happens is when $\exists \epsilon>0$ such that $p(u)>0$ for
all $|u|<a/2+\epsilon$.
\end{enumerate}
We will see that in the first case we obtain a rapidly decreasing
$W(F)$ even though there is no constraint toward Gaussianity in the large 
$N$ limit, while
in the second case we have a power law tailed $W(F)$ similarly to that
of the three-dimensional Holtzmark distribution \cite{chandra}.

\section{Detailed analysis of Eq.~(\ref{eq7})}

Let us analyze the single factor of Eq.~(\ref{eq7}) which, as
aforementioned, is the FT of the conditional PDF $g_n(f;x_0)$ of the
force felt by the particle in $X_0=x_0$ due to only the particle in
$X_n$: 
\bea 
&&\int_{-\infty}^{+\infty}dx\, p(x+x_0-na) \exp
\left(ik\frac{x}{|x|^{\alpha+1}}\right)\nonumber\\
&&= \left<\exp
\left(ik\frac{x}{|x|^{\alpha+1}}\right)\right>_{n,x_0}\,,
\label{eq8}
\eea 
where $\left<s(x)\right>_{n,x_0}$ denotes the average of the
function $s(x)$ over the shifted PDF $h_{n,x_0}(x)=p(x+x_0-na)$.
In practice, if we indicate with simply
$\left<s(u)\right>=\int_{-\infty}^{+\infty}du\,p(u)s(u)$,
then we can say that 
\[\left<s(x)\right>_{n,x_0}=\left<s(u+na-x_0)\right>\,.\]
We want to study Eq.~(\ref{eq8b}) in the limit of small $k$.
Similarly to what pointed out in the previous section, the small $k$
behavior of
$\left<\exp\left(ik\frac{x}{|x|^{\alpha+1}}\right)\right>_{n,x_0}$ is
different in the two cases in which, as a consequence of
displacements, the pair of particles initially in $x=0$ and $x=na$
cannot or can be found arbitrarily close to one each other, i.e.,
respectively if the supports of $p(u)$ and $p(u-na)$ have an empty or
a non-zero overlap.

Let us start with the case (i).  If $\exists\,0<u^*<|n|a/2$ such that
$p(u)=0$ for $|u|\ge u^*$, the exponent of $\exp
\left(ik\frac{x}{|x|^{\alpha+1}}\right)$ can take only limited values in
the integral (\ref{eq8}) (i.e., the support of $g_n(f;x_0)$ is
restricted to only a finite interval of values of $f$). 
Note that in
the given hypothesis (i), if $n>0$ the quantity $x$ can take
only strictly positive values, while if $n<0$ it takes only strictly
negative values. In this case, if $n>0$, we can write: 
\bea 
&&\left<\exp
\left(ik\frac{x}{|x|^{\alpha+1}}\right)\right>_{n,x_0}\nonumber\\
&&= \int_{-\infty}^{+\infty}dx\,
p(x+x_0-na) \sum_{m=0}^{+\infty}\frac{(ikx^{-\alpha})^m}{m!}\nonumber\\
&&=\sum_{m=0}^{+\infty}\frac{(ik)^m}{m!}\left<x^{-\alpha m}\right>_{n,x_0}
\nonumber\\
&&=\sum_{m=0}^{+\infty}\frac{(ik)^m}{m!}\left<(u+na-x_0)^{-\alpha
m}\right>\,,
\label{eq9}
\eea
where 
\[\left<(u+na-x_0)^{-\alpha m}\right>=
\int_{-u^*}^{u^*}du\, p(u) (u-x_0+na)^{-\alpha m}\,.\] 
If instead $n<0$, Eq.~(\ref{eq9}) becomes
\bea 
&&\left<\exp
\left(ik\frac{x}{|x|^{\alpha+1}}\right)\right>_{n,x_0}
=\sum_{m=0}^{+\infty}\frac{(-ik)^m}{m!}\left<(-x)^{-\alpha m}\right>_{n,x_0}
\nonumber\\
&&=\sum_{m=0}^{+\infty}\frac{(-ik)^m}{m!}\left<(-u-na+x_0)^{-\alpha
m}\right>\,,
\label{eq9b}
\eea Note that, as $p(u)=p(-u)$, we have
$\left<(-u-na+x_0)^{-\alpha m}\right>= \left<(u-na+x_0)^{-\alpha
m}\right> $. In both cases
\[\left|\left<\left(\frac{x}{|x|^{\alpha+1}}\right)^m\right>_{n,x_0}
\right|<(|n|a+2u^*)^{-\alpha m}\] 
for any $m\ge 0$, and therefore the series in Eq.~(\ref{eq9})
absolutely converges.  It is very important to note that, if $u^*<a/2$
(i.e., no pair of particles can be found arbitrarily close to one each 
other)
all the factors in Eq.~(\ref{eq7}) can be represented as a Taylor
series (\ref{eq9}) to all orders $m$. As shown below in more detail, this
implies that when $u^*<a/2$, $W(F)$ has all finite moments and
therefore is rapidly decreasing at large $F$.

In the second case (ii) instead $\exists 0<\epsilon<a/2$ such that,
at least $\forall u$ satisfying $|n|a/2-\epsilon<|u|< |n|a/2+\epsilon$, 
we have $p(u)>0$. In this case the
particle initially at $\pm na$ and the particle initially at the
origin can be found arbitrarily close. This implies that the quantity $x$
(i.e., $u-x_0+na$) in (\ref{eq8}) is permitted to take arbitrary small
values up to zero, and therefore in the last expression of Eq.~(\ref{eq9})
there would be an infinite number of diverging terms of the last
series. In other words the Taylor series sum in the second expression
of Eq.~(\ref{eq9}) cannot be exchanged with the average operation
$\left<...\right>_{n,x_0}$, and we expect a singular part in the small
$k$ expansion of the average
$\left<\exp\left(ik\frac{x}{|x|^{\alpha+1}} \right)\right>_{n,x_0}$.
In order to find it, we rewrite it as in Eq.~(\ref{eq8b}):
\bea
&&\left<\exp\left(ik\frac{x}{|x|^{\alpha+1}}\right)\right>_{n;x_0}\equiv
\int_{-\infty}^{+\infty}df\,g_n(f;x_0)e^{ikf}\nonumber\\
&&=\int_{-\infty}^{+\infty}dx\, p(x+x_0-na) \exp
\left(ik\frac{x}{|x|^{\alpha+1}}\right)
\nonumber
\eea
Note that in this case, differently from the previous 
one, the support of $g_n(f;x_0)$ includes arbitrarily large values of
$|f|$ for which, using Eq.~(\ref{gn}), we have 
\[g_n(f;x_0)=\frac{p(x_0-na)}{\alpha}|f|^{-(\alpha+1)/\alpha}+
o(|f|^{-(\alpha+1)/\alpha}).\] 

Let us call $M=[\alpha^{-1}]$ the integer part of $\alpha^{-1}$. 
By using the results presented in Appendix I we can conclude that 
\bea
\label{eq10b}
&&\left<\exp\left(ik\frac{x}{|x|^{\alpha+1}}\right)\right>_{n;x_0}\\
&&=\sum_{m=0}^{M}\frac{(ik)^m}{m!}
\left<(u+na-x_0)^{-\alpha m}\right>+S_n(k;x_0)\,,\nonumber
\eea
where $S_n(k;x_0)$ contains all the terms of order higher than $M$,
including the singular part of the small $k$ expansion of 
$\left<\exp\left(ik\frac{x}{|x|^{\alpha+1}}\right)\right>_{n;x_0}$
which is of order $1/\alpha$ in $k$. By using Eq.~(\ref{app-I-last}),
we can finally write:
\bea
\label{eq10c}
&&S_n(k;x_0)=\\
&&\left\{
\begin{array}{ll}
\frac{(-1)^{(M+1)/2}\pi p(x_0-na)}{\alpha\Gamma[(\alpha+1)/\alpha]
\cos\left(\frac{\alpha^{-1}-M}{2}\pi\right)}k^{1/\alpha}+o(k^{1/\alpha})
&\mbox{for odd }M \\
\\
\frac{(-1)^{M/2}\pi p(x_0-na)}{\alpha\Gamma[(\alpha+1)/\alpha]
\sin\left(\frac{\alpha^{-1}-M}{2}\pi\right)}k^{1/\alpha}+o(k^{1/\alpha})
&\mbox{for even }M
\end{array}
\right.
\nonumber
\eea
Here, for simplicity, we have excluded the case in which exactly
$M=1/\alpha$ for which we have logarithmic corrections in $k$ 
to the above equations.

\section{Finding $W(F)$}

At this point we can go further and classify the possible behaviors of 
$P(F;x_0)$ and $W(F)$.

Basically we again distinguish the following two cases:
\begin{enumerate}
\item $\exists \epsilon>0$ such that $\forall |u|>a/2-\epsilon$
one has $p(u)=0$;
\item $\exists 0<\epsilon<a/2$ such that, at least $\forall u$ 
satisfying $a/2-\epsilon<|u|\le a/2+\epsilon$, 
$p(u)>0$.
\end{enumerate}

\subsection{Case 1: fast decreasing $W(F)$}

In this case the system satisfies the NOC and
all the factors in Eq.~(\ref{eq7}) can be expanded in the
Taylor series (\ref{eq9}) and (\ref{eq9b}) for all different $n$.

We can then write
\bea
\label{eq13}
\hat P(k;x)&\equiv&\int_{-\infty}^{+\infty}dF\,P(F;x)\,e^{ikF}\\
&=&\prod_{n=1}^N \left<e^{i\frac{k}{(u-x+na)^\alpha}}\right>
\left<e^{-i\frac{k}{(u+x+na)^\alpha}}\right>\nonumber\\
&=& \prod_{n=1}^N\left[
\sum_{m=0}^{+\infty}\frac{(ik)^m}{m!}\left<(u-x+na)^{-\alpha m}\right>
\right.\nonumber\\
&&
\left.\sum_{l=0}^{+\infty}\frac{(-ik)^l}{l!}\left<(u+x+na)^{-\alpha
l}\right>\right]\,,
\nonumber 
\eea 
where in the average $\left<...\right>$ over the displacement $u$ 
we have used the symmetry property $p(u)=p(-u)$. 
Note that from Eq.~(\ref{eq13}), we have $\hat P(k;-x)=\hat P^\dag(k;x)$,
where $A^\dag$ indicates the complex conjugate of $A$.

By calling again $u^*<a/2$ the maximal permitted displacement for each
particle (i.e., the support of $p(u)$ is included in $[-u^*,u^*]$), we
can find $\hat W(k)={\cal F}[W(F)]$ by simply calculating the
following average 
\be 
\hat W(k)=\int_{-u^*}^{u^*}dx\,p(x)\hat P(k;x)
\,.
\label{eq14}
\ee 
It is simple to verify that, as $\hat P(k;-x)=\hat P^\dag(k;x)$
and $p(u)=p(-u)$, the function $\hat W(k)$ is real and $\hat W(k)=\hat
W(-k)$.  The Taylor expansion in $k$ of $\hat W(k)$ is obtainable from
Eqs.(\ref{eq13}) and (\ref{eq14}). Since it is a real function and is
the characteristic function only even powers of $k$ are present.  In
particular the coefficient of the $k^2$ term is $-\overline{{\cal
F}^2}/2$ where
\[\overline{h({\cal F})}=\int_{-\infty}^{+\infty}dF\,W(F)h(F)\,.\]
Actually, rigorously speaking, we should show that all the
coefficients of the Taylor expansion of $\hat W(k)$ are convergent to
finite values in the limit $N\rightarrow \infty$.  It is simple to
show it by expanding the terms $\left<(u\pm x_0+na)^{-\alpha
m}\right>$ of Eq.~(\ref{eq13}) in Taylor series of $(u\pm x_0)/na$ for
$n\ge 1$ which is justified by the fact that in the given hypothesis
$|u|+|x_0|\le 2u^*< a$, and considering that
\[
\sum_{m=0}^{+\infty}\frac{(ik)^m}{m!}(na)^{-\alpha m}\times
\sum_{l=0}^{+\infty}\frac{(-ik)^l}{l!}(na)^{-\alpha l}=1\,,\;\;
\forall n\ge 1\,.
\]
Therefore we conclude that we can write $\hat W(k)$ in the following
form:
\be
\hat W(k)=\sum_{n=0}^{+\infty} (-1)^n \frac{\overline{{\cal F}^{2n}}}{(2n)!} 
k^{2n}\,.
\nonumber
\ee
It is simple to see that 
\bea
\label{eq16}
&&\frac{\overline{{\cal F}^2}}{2}=
\sum_{n=1}^N\left\{\left<\left<(u-x+na)^{-2\alpha}
\right>_u\right>_x\right.\\
&&\left.-\left<\left<(u-x+na)^{-\alpha}\right>_u
\left<(u+x+na)^{-\alpha}\right>_u\right>_x
\right\}\nonumber\\
&&+\sum_{n<l}^{1,N}\left<\left[\left<(u-x+na)^{-\alpha}\right>_u-
\left<(u+x+na)^{-\alpha}\right>_u\right]\right.\nonumber\\
&&\left.\times
\left[\left<(u-x+la)^{-\alpha}\right>_u-
\left<(u+x+la)^{-\alpha}\right>_u\right]\right>_x\,,\nonumber
\eea
where for clarity we have redefined
\[\left\{
\begin{array}{l}
\left<a(u)\right>_u=\int_{-u^*}^{u^*}du\,a(u)p(u)\\
\left<a(x)\right>_x=\int_{-u^*}^{u^*}dx\,a(x)p(x)\\
\left<\left<b(u,x)\right>_u\right>_x=\int_{-u^*}^{u^*}\int_{-u^*}^{u^*}
du\,dx\,b(u,x)p(u)p(x)\,.
\end{array}
\right.\] 
It is matter of simple algebra to show that, for
$p(u)=p(-u)$, Eq.~(\ref{eq16}) can be rewritten as \bea
&&\frac{\overline{{\cal
F}^2}}{2}\!=\!\sum_{n=1}^N\!\left<\left<(u-x+na)^{-2\alpha}
\right>_u-\left<(u-x+na)^{-\alpha}\right>_u^2\right>_x\nonumber\\
&&+\sum_{n,l}^{1,N}\left<\left<(u-x+na)^{-\alpha}\right>_u[
\left<(u-x+la)^{-\alpha}\right>_u\right.\nonumber\\
&&\left.-\left<(u+x+la)^{-\alpha}\right>_u] \right>_x
\label{eq16b}
\eea 
We see that the force variance is composed of
two different contributions: the former, given by the first sum in
Eq.~(\ref{eq16b}), is mainly due to the fluctuations in the
displacements $u$ of all the sources of the force (in this term the
average over $x$ is only a smoothing operation), while the latter,
given by the second sum, is determined basically by the fluctuations
created by the stochastic displacement $x$ of the particle initially
in the origin on which we evaluate the force (in this term is the
averages over $u$ to play a role of simple smoothing).

It is interesting and useful in applications to calculate all the
above expressions by evaluating all the terms in the sums in
Eq.~(\ref{eq16b}) to the second order in $(u\pm x)/na$. In order to do
this, we use the following second order Taylor expansion for $B\ll A$:
\bea
&&(A+B)^{-\gamma}=A^{-\gamma}\left(1+\frac{B}{A}\right)^{-\gamma}
\nonumber\\
&&=
A^{-\gamma}\left[1-\gamma\frac{B}{A}+\frac{\gamma(\gamma+1)}{2}
\left(\frac{B}{A}\right)^2+o\left(\frac{B}{A}\right)^2\right]\,.
\nonumber
\eea
From this, substituting respectively $A$ with $na$ and $B$ with $u\pm
x$, we have that
\bea
&&(u\pm x+na)^{-\gamma}\nonumber\\
&&\simeq (na)^{-\gamma}\left[1-\gamma\frac{u\pm x}{na}+
\frac{\gamma(\gamma+1)}{2}\left(\frac{u\pm x}{na}\right)^2\right]\,.
\nonumber
\eea
Moreover we have that $\left<\left<(u\pm
x)^{2n+1}\right>_u\right>_x=0$ for any integer $n$ due to the symmetry
$p(u)=p(-u)$, while we have that $\left<\left<(u\pm
x)^2\right>_u\right>_x=2\sigma^2$ where
$\sigma^2=\left<u^2\right>_u=\int_{-u^*}^{u^*}du\,u^2p(u)$ is the
variance of the single displacement.  Therefore we can write
\[\left<\left<(u\pm x+na)^{-\gamma}\right>_u\right>_x
\simeq (na)^{-\gamma}\left[1+\gamma(\gamma+1)
\frac{\sigma^2}{(na)^2}\right]\,,\]
and
\[\left<\left<(u\pm x+na)^{-\alpha}\right>_u^2\right>_x\!\simeq\! 
(na)^{-2\alpha}\left[1+\alpha(3\alpha+2)\frac{\sigma^2}{(na)^2}\!
\right]\,.\]
Henceforth
\bea
&&\left<\left<(u\pm x+na)^{-2\alpha}\right>_u-
\left<(u\pm x+na)^{-\alpha}\right>_u^2\right>_x\nonumber\\
&&=\frac{\alpha^2\sigma^2}{(na)^{2(\alpha+1)}}\,.
\nonumber
\eea
Moreover
\bea 
&&\left<\left<(u-x+na)^{-\alpha}\right>_u\left<(u\pm x+la)^{-\alpha}
\right>_u\right>_x\nonumber\\
&&\simeq 
(na)^{-\alpha}(la)^{-\alpha}\nonumber\\
&&\times\left[1\mp \frac{\alpha^2\sigma^2}{(la)(na)}+
\alpha(\alpha+1)\sigma^2\left(\frac{1}{(na)^2}+\frac{1}{(la)^2}\right)
\right]\,;\nonumber
\eea
from which
\bea 
&&\left<\left<(u-x+na)^{-\alpha}\right>_u[\left<(u- x+la)^{-\alpha}
\right>_u\right.\\
&&\left.-\left<(u+x+la)^{-\alpha}
\right>_u]\right>_x\simeq
\frac{2\alpha^2\sigma^2}{(la)^{\alpha+1}(na)^{\alpha+1}}\,.\nonumber
\eea
Using all these results in all the terms of Eq.~(\ref{eq16b}), we obtain
\be
\frac{\overline{{\cal F}^2}}{2}\simeq\alpha^2\sigma^2\left\{
\sum_{n=1}^N\frac{1}{(na)^{2(\alpha+1)}}+2\left[
\sum_{n=1}^N\frac{1}{(na)^{\alpha+1}}\right]^2\right\}
\label{eq17}
\ee
It is simple to verify that both sums in Eq.~(\ref{eq17}) are
converging for $N\rightarrow +\infty$ for all $\alpha>0$, for which we can then
rewrite
\be
\frac{\overline{F^2}}{2}\simeq\frac{\alpha^2\sigma^2}{a^{2(\alpha+1)}}\left[
\zeta(2\alpha+2)+2
\zeta^2(\alpha+1)\right]\,,
\label{eq18}
\ee where $\zeta(t)$ is the {\em Riemann zeta function} (note that for
$t\rightarrow 1^+$ wee have $\zeta(t)\simeq 1/(t-1)$).  Again the
first term is due to the fluctuations in the position of the sources,
while the second one is due to the fluctuations in the position of the
particle on which we are calculating the force.  In particular In
Eq.~(\ref{eq17}) the generic term of the first sum give the relative
weight of the $n^{th}$ nearest neighbor particles in determining the
force on the particle in $X_0$.  At last we can say that, in the case
of displacements limited within a box well contained in a unitary
cell around the initial lattice position, 
we can approximate $W(F)$ with a Gaussian PDF with zero mean and
variance given by Eq.~(\ref{eq18}).  However, as already pointed out,
there is no constraint, in the limit $N\rightarrow\infty$, toward
rigorous Gaussianity and non-Gaussian corrections are in general
present.

\subsection{Case 2: power law tailed $W(F)$}

As shown above, this is the case in which the OC is satisfied,
i.e., particles are permitted to
jump out of their initial lattice positions beyond the limit of the
related unitary cell in such a way to be found arbitrarily close to
some other particle.  Note that this is always the case when the
support of $p(u)$ is unlimited, i.e., if $p(u)>0$, $\forall u \in
I\!\!R$.  However the same kind of $W(F)$ is also obtained if $\exists
u^*>a/2$ such that $p(u)>0$, $\forall u\in [-u^*,u^*]$ and zero
outside.  The difference between these two sub-cases is only in the
amplitude of the power law tail of $W(F)$ but not in its exponent.  In
general, if the particle initially at the lattice site $x=na$ is
permitted, through displacements, to be found arbitrarily close to the
particle initially at $x=0$, it will contribute to the product
(\ref{eq7}) through a factor of the type (\ref{eq10b}). If instead
this is not permitted, it will contribute to (\ref{eq7}) through a
factor of the form (\ref{eq9}) or (\ref{eq9b}) depending respectively
on whether $n>0$ or $n<0$. In any case if $M=[1/\alpha]$, in order to
find the main terms of the small $k$ expansion of $\hat W(k)$ (so to
determine the large $F$ tail of $W(F)$), it is sufficient to truncate
all the small $k$ expansion of the different factors in Eq.~(\ref{eq7})
at most to the order $M+1$. For the sake of simplicity, let us limit
the discussion to the case in which strictly $\alpha M<1$ in such a
way to exclude logarithmic corrections in $k$.  We can write 
\bea 
\label{eq19}
&&\hat P(k;x_0)\\ &&\simeq\prod_{n}^{\tiny (\mbox{OC})}
\left[\sum_{m=0}^M\frac{(ik)^m}{m!}
\left<\left(\frac{x}{|x|^{\alpha+1}}
\right)^m\right>_{n,x_0}\!\!+A(n,x_0,\alpha)k^{1/\alpha}\right]\nonumber\\
&&\times\prod_{l}^{(\tiny
\mbox{NOC})}\left[\sum_{m=0}^{M+1}\frac{(ik)^m}{m!}
\left<\left(\frac{x}{|x|^{\alpha+1}}\right)^m\right>_{l,x_0}\right]\,,
\nonumber \eea where $A(n,x_0,\alpha)$ is the coefficient of the term
$\sim k^{1/\alpha}$ in Eq.~(\ref{eq10c}), and the first product on $n$
is on the particles in $X_n$ with $n\ne 0$ which can be found
arbitrarily near to the particle in $X_0$ (i.e., satisfying the OC
with respect to the particle in $X_0$), while the product on $l$ is on
the particles in $X_l$ with $l\ne 0$ which have a positive minimal
distance to the same particle (i.e., satisfying the NOC with respect
to the particle in $X_0$).  If $p(u)>0$ $\forall u\in I\!\!R$ all the
system particles with $n\ne 0$ are included in the first product.  If
instead $p(u)>0$ for $u\in [-u^*,u^*]$ with $u^*>a/2$ and zero
outside, the first product include only contributions from the
particles with $-2u^*/a<n<2u^*/a$ and $n\ne 0$, while the others are
included in the second product. The large $F$ behavior of $P(F;x_0)$
and consequently of $W(F)$ is completely determined by the singular
term of order $k^{1/\alpha}$ in the small $k$ expansion of $\hat
P(k;x_0)$.  It is simple to see that up to the order $k^{1/\alpha}$
\be \hat P(k;x_0)=\sum_{m=0}^{M}c_m(x_0) k^m
+c_{1/\alpha}(x_0)k^{1/\alpha}\,, \nonumber \ee where the the
coefficients $c_m(x_0)$ can be deduced by counting from
Eq.~(\ref{eq19}) (in particular $c_m(x_0)=i^m \widetilde{{\cal
F}^m}(x_0)/m!$, where $\widetilde{{\cal
F}^m}(x_0)=\int_{-\infty}^{+\infty}dF\,P(F;x_0)F^m$ is the $m^{th}$
moment of $P(F;x_0)$ and $m\le M$) and
\[c_{1/\alpha}=\sum_{-2u^*/a<n<2u^*/a}^{n\ne 0}
A(n;x_0)\,,\]
where the formula includes also the case $u^*\rightarrow\infty$.
The small $k$ expansion up to the order $k^{1/\alpha}$ of $W(F)$ will
be consequently
\be
\hat W(k)\simeq \sum_{m=0}^M b_m k^m +b_{1/\alpha}k^{1/\alpha}\,,
\nonumber
\ee
where
\bea
b_m&=&\int_{-\infty}^{+\infty}dx_0\, p(x_0)c_m(x_0) =\frac{i^m}{m!}
\overline{{\cal F}^m}\mbox{ with }m\le M\nonumber \\
b_{1/\alpha}&=& \int_{-\infty}^{+\infty}dx_0\, p(x_0)c_{1/\alpha}(x_0)\,,
\nonumber
\eea 
where $\overline{{\cal F}^m}=\int_{-\infty}^{+\infty}dF\,W(F)F^m=
\int_{-\infty}^{+\infty}dx_0\,p(x_0)\widetilde{{\cal F}^m}(x_0)
\equiv \left<\widetilde{{\cal F}^m}(x_0)\right>$.  It is possible to
evaluate explicitly $b_{1/\alpha}$ by using Eq.~(\ref{eq10c}).  

We are now in the situation to connect the singular term
$b_{1/\alpha}k^{1/\alpha}$ of the small $k$ expansion of $\hat W(k)$
to the large $F$ tail of $W(F)$ by using directly the arguments in
Appendix I.  This gives simply 
\be 
W(F)\simeq B\,F^{-1-1/\alpha}
\nonumber
\ee
with
\be
B= \frac{1}{\alpha}\int_{-\infty}^{+\infty}dx_0\, p(x_0)
\sum_{-2u^*/a<n<2u^*/a}^{n\ne 0}p(x_0-na)\,.
\label{eq23}
\ee
Note that if the support of $p(x_0-na)$ is much larger than $a$ and
$p(u)$ is smooth (i.e., approximately constant) on the scale $a$, we
can approximate Eq.~(\ref{eq23}) with
\be 
B=\frac{1-p(0)\,a}{\alpha a}\,.
\nonumber
\ee
Note that this last approximated expression is not dependent on the
details of $p(u)$ for $u\ne 0$.
Finally, we can observe that we have obtained a power law 
tailed $W(F)$ characterized by the same exponent of the case of a
homogeneous Poisson particle distribution presented in Appendix II.
The only differences are the two following:
\begin{itemize}
\item The amplitude of this power law tail is reduced in the shuffled
lattice with respect to that of the Poisson particle distribution,
given by Eq.~(\ref{app2-10}), of
a factor
\[\int_{-\infty}^{+\infty}dx_0\, p(x_0)
\sum_{-2u^*/a<n<2u^*/a}^{n\ne 0}a\,p(x_0-na)\simeq 1-p(0)\,a\,.\]
\item In the shuffled lattice we have this power law tail for each 
$\alpha>0$, while in the Poisson case the problem is not well defined 
for $\alpha\le 1/2$ (see Appendix II).
\end{itemize}


\section{Conclusions}

We have presented a detailed study of the PDF $W(F)$ of the stochastic
force ${\cal F}$ generated by a randomly perturbed lattice of
sources of a scale invariant attractive pair interaction field 
$f(x)=-Cx/|x|^{\alpha+1}$ with $\alpha>0$ at distance $x$ 
from the source.

In general we distinguish two cases: 
\begin{enumerate}
\item The NOC is satisfied and no pair of particles can be
found at an arbitrarily small reciprocal distance; 
\item The OC is satisfied and it exists at
least one of such pairs of particles.  
\end{enumerate}
In the first case we have a
fast decreasing $W(F)$ similar to a Gaussian PDF at large $F$, even
though no constraint toward an exact Gaussian central limit theorem is
found. In the second case a power law tailed $W(F)$ is found.  The
unique exponent of such power law is directly related to the pair
interaction exponent $\alpha$, while its amplitude depends also on the
lattice spacing $a$ (with respect to the unit distance through which
we measure $x$ in $f(x)$) and in general on the shape of the
perturbations PDF $p(u)$. In particular in this case $W(F)$ has a power law 
tail with the same exponent as the stable L\'evy distribution found 
in the Poisson case (see Appendix II) but with a reduced
amplitude, even though, analogously to the case (i), no constraint has
been found toward the stable L\'evy distribution.

Some further general considerations have can now be done:
\begin{itemize}
\item In the case in which the probability of finding arbitrarily
close to one each other, the large $F$ behavior of $W(F)$ is basically
determined by the small $x$ behavior of $f(x)$ and not at all by the
the fact if it is long range or not.  Therefore
if we considered a fast decreasing $f(x)$ but with the same divergence
in $x=0$ we would have deduced the same conclusions about the exponent
of the large $F$ tail of $W(F)$
\footnote{It is possible to show \cite{andrea-corr} instead that
the large $x$ behavior of the pair interaction $f(x)$ determines the
large $|x-y|$ behavior of the field-field correlation function
$\left<{\cal F}(x){\cal F}(y)\right>$.}.
\item For this reason, even if we consider a lattice perturbed by {\em
correlated} displacements, we expect to obtain the partition into the
two cases (i) and (ii) above considered depending on the possibility
or not to find pair of particles arbitrarily close.  
\item Actually different cases for the large $F$ tail of $W(F)$
between the Gaussian-like ``fast decreasing'' and L\'evy-like ``power
law'' tailed PDF with exponent $\beta=(\alpha+1)/\alpha$ are possible
in very particular cases.  These cases correspond to the choice of
$p(u)$ such that $p(u)>0$ exactly for $u\in [-a/2,a/2]$ and zero
outside.  By changing the limit behaviors of $p(u)$ when $u\rightarrow
\pm a/2$ we can obtain different large $F$ behaviors of $W(F)$.
In particular if $p(a/2)>0$ and finite (we consider $p(u)=p(-u)$) we have the
same case as (ii) described above with $\beta= (\alpha+1)/\alpha$.  If
instead $p(a/2)=0$, depending on the behavior of $p(u)$ for
$u\rightarrow (a/2)^-$ we will have different values of $\beta$ but in
general larger than $(\alpha+1)/\alpha$. If, finally, $p(a/2)=+\infty$
(in such a way that $p(u)$ remains anyway integrable) in general we obtain
$\beta$ smaller than $(\alpha+1)/\alpha$ (but always $>1$ so that
$W(F)$ remains integrable).
\end{itemize} 

\section*{Appendix I: Fourier transform of power law tailed PDFs}

We are interested in the small $k$ behavior of the characteristic function
$\hat f(k)$ of a given power law tailed PDF $f(x)$ which for large $|x|$ 
behaves as $A|x|^{-\alpha}$ with $\alpha >1$.
Let us call $[\alpha]=n\ge 1$ the integer part of $\alpha$. 
In this hypothesis $\hat f(k)$ has a regular Taylor expansion up to the 
order $n-1$ followed by a singular term proportional to $k^{\alpha-1}$:
\be
\hat f(k)\equiv \int_{-\infty}^{+\infty}dx\,f(x)e^{ikx}=
\sum_{m=0}^{n-1}\frac{(ik)^m}{m!}\overline{x^m}+\hat f_s(k)\,,
\label{app1-0}
\ee 
where $\overline{x^m}=\int_{-\infty}^{+\infty}dx\,x^m p(x)$ and 
$\hat f_s(k)$ contains the singular part of $\hat f(k)$ and at small $k$ is
an infinitesimal of order $\alpha-1$ in $k$ (if
$\alpha$ is an integer it contains also logarithmic corrections).

Now we show that effectively at sufficiently small $k$, $\hat f_s(k)
\sim B k^{\alpha-1}$ (where now $a(k) \sim b(k)$ means that
$\lim_{k\rightarrow 0}[a(k)/b(k)]=1$) giving an explicit expression for
$B$ as a function of both the amplitude $A$ and the exponent $\alpha$.

First of all, let us study the case of a function $h(x)$ that can be
written as
\be
h(x)=B|x|^{-\beta}+h_0(x)\,,
\label{app1-0b}
\ee where $B>0$, $0<\beta<1$ and $h_0(x)$ is a smooth function,
integrable in $x=0$ and such that $x^{\beta}h_0(x)\rightarrow 0$ for
$|x|\rightarrow\infty$.  
This means that $h(x)$ presents an even power law tail. 
In this case the small $k$ behavior of $\hat
h(k)= \int_{-\infty}^{+\infty}dx\,h(x)e^{ikx}$ is completely
determined by the Fourier transform of $B|x|^{-\beta}$, i.e.: 
\[ \hat
h(k)\sim B\int_{-\infty}^{+\infty}dx |x|^{-\beta} e^{ikx}\,\]
In order to perform this Fourier transform we introduce the integral 
representation:
\be
|x|^{-\beta}=\frac{1}{\Gamma(\beta)}\int_0^\infty dz\,z^{\beta-1}e^{-|x|z}\,,
\label{app1-2}
\ee
where $\Gamma(\beta)$ is the Euler Gamma function. Using Eq.~(\ref{app1-2})
we can write:
\bea
&&\int_{-\infty}^{+\infty}\!dx |x|^{-\beta} e^{ikx}=
\frac{1}{\Gamma(\beta)}\int_0^\infty\! dz\,z^{\beta-1}
\int_{-\infty}^{+\infty}\!dx e^{ikx-|x|z}\nonumber\\
&&=\frac{2k^{\beta-1}}{\Gamma(\beta)}
\int_0^\infty dq\frac{q^\beta}{1+q^2}\,.
\nonumber
\eea
Using the general relation valid for $0<\beta<1$
\be
\int_0^\infty dq\frac{q^\beta}{1+q^2}=\frac{\pi}{2\cos 
\left(\frac{\beta}{2}\pi\right)}\,,
\label{app1-3b}
\ee
we can conclude
\be
\hat h(k)=\frac{B\pi}{\Gamma(\beta)\cos\left(\frac{\beta}
{2}\pi\right)}k^{\beta-1}+o(k^{\beta-1})\,.
\label{app1-4}
\ee
If instead of using the integral representation (\ref{app1-2}) one used
\[
|x|^{-\beta}=\frac{1}{\Gamma\left(\frac{\beta}{2}\right)}
\int_0^\infty dz\,z^{\beta/2-1}e^{-x^2z}\,,
\]
one should obtain the alternative expression containing only Gamma
functions:
\[
\hat h(k)=\frac{B\sqrt{\pi}\Gamma\left(\frac{1-\beta}{2}\right)}{2^\beta
\Gamma\left(\frac{\beta}{2}\right)}k^{\beta-1}
+o(k^{\beta-1})\,.
\]
Note that the coefficient of the term $k^{\beta-1}$ is real and positive.
Another important case is when the function $h(x)$ has an odd non-integrable 
power law tail, i.e.: 
\be
h(x)=B|x|^{-\beta}[2\theta(x)-1]+h_0(x)\,,
\label{app1-5}
\ee
where $\theta(x)$ is the usual Heaviside step function, $B>0$, $0<\beta<1$,
and $h_0(x)$ the same of Eq.~(\ref{app1-0b}).  By using the same integral
transformation leading to Eq.~(\ref{app1-4}), we in this case we obtain:
\[
\hat h(k)=i\frac{B\pi}{\Gamma(\beta)\sin\left(\frac{\beta}
{2}\pi\right)}k^{\beta-1}+o(k^{\beta-1})\,.
\]
At this point we can go back to the problem of finding the dominant
small $k$ contribution of the term $\hat f_s(k)$ in Eq.~(\ref{app1-0}) for 
the PDF $f(x)$ decaying at large $|x|$ as $A|x|^{-\alpha}$.
Note that now we cannot apply directly the argument we have used
for the above function $h(x)$.
In fact if, from one side, also in this case we can write
\[
f(x)=A|x|^{-\alpha}+f_0(x)
\]
with $|x|^{\alpha}f_0(x)\rightarrow 0$ for $|x|\rightarrow \infty$,
from the other side $\alpha>1$ (for definiteness of probability)
and $f_0(x)$ contains a non integrable singularity at $x=0$ such that
to cancel the non integrable contribution of the $A|x|^{-\alpha}$ term
at small $x$.
In order to circumvent this difficulty we introduce the function
\[g(x)=x^n f(x)\,,\]
where $n$ is the integer part of $\alpha$.  In this way $g(x)$ is
similar to the function $h(x)$ of Eq.~(\ref{app1-0b}) if $n$ is even
and to the $h(x)$ of Eq.~(\ref{app1-5}) if $n$ is odd.
Therefore, by defining as usual 
$\hat g(k)=\int_{-\infty}^{+\infty}dx\,g(x)e^{ikx}$, we can say that
\[\hat g(k)=
\frac{A\pi}{\Gamma(\alpha-n)\cos\left(\frac{\alpha-n}
{2}\pi\right)}k^{\alpha-n-1}+o(k^{\alpha-n-1})\]
if $n$ is even, and
\[\hat g(k)=i
\frac{A\pi}{\Gamma(\alpha-n)\sin\left(\frac{\alpha-n}
{2}\pi\right)}k^{\alpha-n-1}+o(k^{\alpha-n-1})\]
if $n$ is odd.
Now in order to find the singular part $\hat f_s(k)$ of $\hat f(k)$
it is sufficient to integrate $n$ times $\hat g(k)$ (the integration 
constants giving rise to the finite moments terms of $\hat f(k)$ in 
Eq.~(\ref{app1-0})).
In this way we obtain
\bea
\label{app-I-last}
&&\hat f_s(k)\\
&&=\left\{
\begin{array}{ll}
\frac{(-1)^{n/2}A\pi}{\Gamma(\alpha)\cos\left(\frac{\alpha-n}
{2}\pi\right)}k^{\alpha-1}+o(k^{\alpha-1})&\mbox{for even }n\\
\\
\frac{(-1)^{(n+1)/2}A\pi}{\Gamma(\alpha)\sin\left(\frac{\alpha-n}
{2}\pi\right)}k^{\alpha-1}+o(k^{\alpha-1})&\mbox{for odd }n\,,
\end{array}
\right.\nonumber
\eea
where we have used the following property of Gamma function: 
$\Gamma(x+1)=x\Gamma(x)$, hence $[(\alpha-1)...(\alpha-n)\Gamma(\alpha-n)]=
\Gamma(\alpha)$.
Note that in both case the coefficient of the term $k^{\alpha-1}$ is
real.

\section*{Appendix II: Force PDF in a Poisson particle distribution}

Let us consider the case in which the particles are distributed on the
line interval $(-L/2,L/2]$ of length $L$ following a spatially
stationary Poisson process with average density $\rho_0>0$.  We want
to know the PDF $W_P(F)$ of the field
\be
F=\sum_{i=1}^N \frac{x_i}{|x_i|^{\alpha+1}}
\label{app2-0}
\ee generated at the origin of the space by all the $N$ system
particles (we can consider $N=\rho_0L$ as the fluctuations of order
$\sqrt{\rho_0L}$, due to the Poisson statistics, are completely
unimportant for this problem in the large $L$ limit).  We will follow
the procedure to find $W_P(F)$ in three dimensions used by Chandrasekhar in
\cite{chandra} for the gravitational force.  Note that, as the
positions of different particles are uncorrelated, the joint PDF
$p(x_1,...,x_N)$ of the positions of the $N$ system particles is
simply:
\[p_N(x_1,...,x_N)=\prod_{i=1}^{N}p(x_i)\,,\]
where $p(x_i)=1/L$.  Therefore we can write
\be
W_P(F)=\int\int_{-L/2}^{L/2}\left[\prod_{i=1}^{N}
\frac{dx_i}{L}\right]\delta\left(F-\sum_{j=1}^N 
\frac{x_i}{|x_j|^{\alpha+1}}\right)\,.
\label{app2-1}
\ee
Let us now study the characteristic function
\[\hat W_P(k)={\cal F}[W_P(F)]=\int_{-\infty}^{+\infty}dF\,W_P(F)e^{ikF}
\,.\] 
By taking the FT of Eq.~(\ref{app2-1}) we obtain
\[
\hat W_P(k)=\left[\int_{-L/2}^{L/2}\frac{dx}{L}\exp\left(ik\frac{x}
{|x|^{\alpha+1}}\right) \right]^N
\]
By adding and subtracting 1 inside the square brackets, and taking the
limit $L\rightarrow\infty$ with $N=\rho_0L$ we arrive at the final
expression:
\[
\hat W_P(k)=\exp\left[-\rho_0k^{1/\alpha}\!\int_{-\infty}^{+\infty}\!dt\,
\left(1-\exp\left(i\frac{t}{|t|^{\alpha+1}}\right)\right) \right]\,.
\]
Note that
\bea
&&\int_{-\infty}^{+\infty}\!\!dt
\left(1-\exp\left(i\frac{t}{|t|^{\alpha+1}}\right)\right)
=2\int_{0}^{\infty}\!\!dt
\left(1-\cos(t^{-\alpha})\right)\nonumber\\
&&=\frac{2}{\alpha}\int_{0}^{\infty}
du\,u^{-1-1/\alpha}(1-\cos u)\,,
\nonumber
\eea
where the last passage is due to the change of variable $t^{-\alpha}=u$.
Let us now use, as in the previous appendix, the integral representation:
\[u^{-1-1/\alpha}=\frac{1}{\Gamma\left(\frac{\alpha+1}{\alpha}\right)}
\int_{0}^{\infty}dz\,z^{1/\alpha}e^{-uz}\,.\]
Through this transformation we arrive finally to the relation
\bea
&&\int_{-\infty}^{+\infty}dt\,
\left(1-\exp\left(i\frac{t}{|t|^{\alpha+1}}\right)\right)\nonumber\\
&&=\frac{2}{\alpha\Gamma\left(\frac{\alpha+1}{\alpha}\right)}
\int_{0}^{\infty}dz\,\frac{z^{-1+1/\alpha}}{1+z^2}\,.
\label{app2-5}
\eea
Note that the last integral is diverging for $\alpha\le 1/2$,
indicating that the problem is not well defined for these values of
$\alpha$ as $F$ is not a well defined stochastic quantity.
This means that the sum in Eq.~(\ref{app2-0})
needs a $L$ dependent normalization to become a well defined
stochastic variable.  In fact, differently to the shuffled lattice 
case, where the typical mass fluctuation
on regions of size $R$ is proportional to $R^0$, in the Poisson case
this is due to the fact that such fluctuation is proportional to
$R^{1/2}$.  The field due to the mass fluctuation in a sphere of
radius $R$ on the origin of the sphere is of order $R^{-\alpha}$ for
the shuffled lattice and $R^{-\alpha+1/2}$ in the Poisson point
process.  This explains why for a shuffled lattice the problem is well
defined for any $\alpha > 0$ and not only for $\alpha>1/2$. This also
says that for $\alpha<1/2$, in order to have a well defined stochastic
field also in the Poisson case, we have to divide the field in
Eq.~(\ref{app2-0}) by $L^{-\alpha+1/2}$ where $L$ is the system size.  
The same argument can be used in $d$
dimensions, for which the same mass fluctuations are respectively
proportional to $R^{(d-1)/2}$ in the shuffled lattice, and $R^{d/2}$
in the Poisson case. This says that the problem is well defined,
without $L$ dependent normalization of the field, for $\alpha>(d-1)/2$
for the shuffled lattice and $\alpha>d/2$ for the Poisson case (in the
gravitational case in $d=3$ faced by Chandrasekhar \cite{chandra} we
have $\alpha=2$ and the field is a well defined stochastic quantity
in both cases).

Let us now go back to Eq.~(\ref{app2-5}) for $\alpha>1/2$. By using 
Eq.~(\ref{app1-3b}) we can rewrite it as
\[
\int_{-\infty}^{+\infty}dt\,
\left(1-\exp\left(i\frac{t}{|t|^{\alpha+1}}\right)\right)=
\frac{\pi}{\alpha\Gamma\left(\frac{\alpha+1}{\alpha}\right)\sin\left(
\frac{\pi}{2\alpha}\right)}\,,
\]
where we have also used $\cos[\pi(1/\alpha-1)/2]=\sin[\pi/(2\alpha)]$.
Consequently
\be
\hat W_P(k)=\exp\left[-
\frac{\pi\rho_0}{\alpha\Gamma\left(\frac{\alpha+1}{\alpha}\right)\sin\left(
\frac{\pi}{2\alpha}\right)}k^{1/\alpha}\right]\,,
\label{app2-7}
\ee
which is exactly of the form of the characteristic function of a L\'evy stable
PDF \cite{levy-stable}.
By expanding Eq.~(\ref{app2-7}) to the first non-vanishing order larger
than zero, we have:
\[
\hat W_P(k)=1-
\frac{\pi\rho_0}{\alpha\Gamma\left(\frac{\alpha+1}{\alpha}\right)\sin\left(
\frac{\pi}{2\alpha}\right)}k^{1/\alpha}+o(k^{1/\alpha})\,.
\]
If we now invert this Fourier transform as explained in the preceding
appendix, we can conclude that 
\[
W_P(F)\simeq B F^{-(\alpha+1)/\alpha}\:\mbox{   for large }F\,,
\]
with 
\be
B=\frac{\rho_0}{\alpha}\,.
\label{app2-10}
\ee

\end{document}